\documentclass[aps,twocolumn,raggedbottom,prl,amsmath,showpacs,nobalancelastpage,amssymb,superscriptaddress]{revtex4}
\usepackage{epsfig}
\begin{document}
\title{Correlated tunneling in intramolecular carbon nanotube quantum dots}
\author{M.~Thorwart}
\affiliation{Department of NanoScience, Delft University of
Technology, Lorentzweg 1, 2628 CJ Delft, The Netherlands}
\author{M.~Grifoni}
\affiliation{Department of NanoScience, Delft University of
Technology, Lorentzweg 1, 2628 CJ Delft, The Netherlands}
\author{G. Cuniberti} 
\affiliation{Max Planck Institute for the Physics of Complex Systems, D-01187, Dresden, Germany}
\author{H.~W.~Ch.~Postma}
\altaffiliation[Current address:]{ Condensed Matter Physics 114-36, 
California Institute of Technology, Pasadena, CA 91125} 
\affiliation{Department of NanoScience, Delft University of
Technology, Lorentzweg 1, 2628 CJ Delft, The Netherlands}
\author{C.~Dekker}
\affiliation{Department of NanoScience, Delft University of
Technology, Lorentzweg 1, 2628 CJ Delft, The Netherlands}
\date{\today}
\begin{abstract}
We investigate correlated electronic transport in single-walled carbon nanotubes  with two
intramolecular tunneling barriers. 
We suggest that below a characteristic temperature the  long range  nature of the 
Coulomb interaction becomes crucial  to determine the
temperature dependence of the maximum $G_{\rm max}$ of the conductance peak.
  { Correlated\/} sequential  tunneling
 dominates  transport yielding
the power-law $G_{\rm max}\propto T^{\alpha_{\rm end-end}-1}$,
typical for tunneling
between the ends of two Luttinger liquids.
Our predictions are in agreement with recent measurements.
\end{abstract}
\maketitle
Electronic correlations have been predicted to
dominate the characteristic features in single-walled metallic
carbon nanotubes (SWNTs)  \cite{Egger97,Kane97},
which has recently been observed in experiments
\cite{Yao99,Bockrath99,Postma00}. The one-dimensional nature of the electronic conduction
bands reveals itself in typical Luttinger liquid  \cite{Voit95}, rather
than Fermi liquid behavior.
The  manipulation of  individual nanotubes with
an atomic force microscope permits the creation of
intra-tube buckles
acting as tunneling barriers \cite{Postma00}.
Recently, SWNTs with two intramolecular buckles have been reported to behave
as room-temperature
single-electron transistors \cite{Postma01}.
At low temperatures, the thermal electronic energies are smaller than the level
separation between the discrete energy states of the island and
tunneling occurs via
these discrete levels.
Resonant tunneling in Luttinger liquids has  been investigated theoretically by many
authors \cite{Kane92,Furusaki93,Sassetti95,Furusaki98,Braggio00}.
 The one-dimensional nature of the correlated electrons
 is responsible for
 the differences to the  quantum Coulomb blockade theory  for conventional, e.g., semiconducting
quantum dots  \cite{Beenakker91}.
Varying the gate voltage results in a sequence of conductance peaks.
In the (uncorrelated) sequential tunneling (UST) approximation the temperature dependence of
the maxima of those peaks follows the power-law \cite{Furusaki98}
\begin{equation}
G_{\rm max}\propto T^{\alpha_{\rm end}-1}\;,
\label{conductanceF}
\end{equation}
with $\alpha_{\rm end}$ being the density of states exponent for tunneling into
 the end of a Luttinger liquid.
However, recent experiments \cite{Postma01} suggest a different power-law,
\begin{equation}
G_{\rm max}\propto T^{\alpha_{\rm end-end}-1} \;,
\label{conductance}
\end{equation}
with $\alpha_{\rm end-end}=2\alpha_{\rm end}$.

In this Letter, we
 propose that a novel tunneling mechanism,
{\em correlated sequential tunneling \/}(CST), gives rise to the power law 
(\ref{conductance}). 
 It originates from the finite
range nature of the Coulomb interaction in SWNTs and 
 {\em replaces\/} conventional uncorrelated sequential tunneling. 
 It dominates
resonant transport at low temperatures and strong 
interactions. CST leads to a  renormalization of the intrinsic
linewidth of the resonance,
 which is reflected in an  increase of $G_{\rm max}$ with increasing temperature. In contrast, UST
would predict the opposite behavior \cite{note}. A good agreement with the experimental
results \cite{Postma01} is found.

 We describe  an individual metallic SWNT with two buckles
 by  the Hamiltonian $H=H_0+H_{\rm B}+H_{\rm ext}$.
  Here, $H_0$
 characterizes the one-dimensional homogeneous wire including the
 finite-range electronic interaction.
 Metallic carbon nanotubes possess two gap-less one-dimensional bands
 \cite{Egger97,Kane97} with
 Fermi velocity $v_{\rm F}\simeq 8\times 10^{5}$ m/s \cite{Postma01}, which  dominate
 the low-energy physics.
In the bosonized representation, one finds \cite{Kane97}
\begin{eqnarray}
H_0 &=& \sum_{a =\rho,\sigma,\Delta\sigma,\Delta\rho}
\int dx \frac{\hbar v_{\rm F}}{2}\left[(\partial_x \vartheta_a)^2
+(\partial_x\phi_a)^2\right]
\nonumber\\ &+& \frac{1}{\pi}\int dx\int dx'
\partial_x\vartheta_\rho(x)V(x-x')\partial_{x'}\vartheta_\rho(x')\;,
 \label{luttliq}
\end{eqnarray}
 containing 
  three sectors of neutral excitations ($a=\sigma,\Delta\rho,\Delta\sigma$)
   propagating with   $v_{\rm F}$,
 and one sector of charged excitations ($a=\rho$).
The charge sector is characterized by the nonlinear dispersion relation
 $\omega (k)=v_{\rm F}|k|[1+\hat V(k)/\pi\hbar v_{\rm F}]$,
where  $\hat V(k)$ is the Fourier transform of the screened
Coulomb potential  projected onto the
 $x$ direction $V(x)$ \cite{Cuniberti98}.
 In our approach, the nonlinear 
  dispersion is crucial, since it modifies the low energy properties
 of the dynamics.
The two tunneling barriers
at $x=\pm x_0/2$ are modeled by a scattering potential $U(x)=U_{\rm B}\sum_{\kappa=\mp}
\delta(x+\kappa x_0/2)$ leading to
$H_{\rm B}=\int dx U(x) \rho(x)$. Here, $\rho(x)$ is the total electron density and
backscattering has to be included. 
 Upon introducing
$ N_a  =  \frac{2}{\sqrt{\pi}} \left[ \vartheta_a
\left(\frac{x_0}{2}\right) + \vartheta_a
\left(-\frac{x_0}{2}\right)\right]$ and $ n_a  =
\frac{2}{\sqrt{\pi}} \left[ \vartheta_a \left(\frac{x_0}{2}\right)
- \vartheta_a \left(-\frac{x_0}{2}\right)\right] +
 \frac{4 k_{\rm F} x_0}{\pi}\delta_{a,\rho}$, $H_{\rm B}$
 provides
 an eight-dimensional sinusoidal
potential for the tunneling dynamics associated with the variables $N_a$ and $n_a$
\cite{Kane92}.
Physically, $eN_\rho/2$ counts the unit charges transferred
through the dot, while $-e n_{\rho}$ is  the total charge accumulated on
 the dot. Likewise, $N_\sigma$ and $n_\sigma$
 are related to the difference of  spin fluctuations between
  both leads, and to the deviation from the mean value
  in the dot, respectively. The analogous interpretation for $N_{\Delta\sigma},n_{\Delta\sigma}$
 and $N_{\Delta\rho},n_{\Delta\rho}$ holds.
Finally, the  gate voltage
$V_{\rm G}$ and the bias voltage $V$ 
give rise to the term
$H_{\rm ext}= -e [ (V/2) N_\rho + \lambda V_{\rm G}  n_\rho]$,
where $\lambda=C_{\rm G}/C_{\rm tot}$,  with $C_{\rm G}$ the gate capacitance
  and $C_{\rm tot}$ the  total capacitance  of gate and island.

The relevant observable is the asymptotic 
tunneling current $I=\lim_{t\rightarrow
\infty}\frac{e}{2}\langle \dot{N_\rho}(t) \rangle$, where $\langle
\dots \rangle$ denotes the expectation value determined via the
reduced density matrix (RDM). The latter is obtained by tracing
out exactly the Luttinger modes
 away from the barriers 
 \cite{Weiss99}. The diagonal elements of the RDM are given by
the  conditional probabilities $P({\mathbf N}_{\rm
f},t; {\mathbf N}_{\rm i},t_0)$ of being in the final state
${\mathbf N}_{\rm f}:=(\{N_{a,{\rm f}}\},\{n_{a,{\rm f}}\})$ at
time $t$, having started from the  state ${\mathbf N}_{\rm
i}$
 at time $t_0$. Then, we find
\begin{equation}
I=\frac{e}{2} \lim_{t\rightarrow \infty}  \sum_{{\mathbf N}_{ \rm
f}} N_{\rho,{\rm f}} \dot{P}({\mathbf N}_{\rm f},t; {\mathbf
N}_{\rm i},t_0):=e(\Gamma^{\rm f}-\Gamma^{\rm b}) \, ,
\label{current}
\end{equation}
with $\Gamma^{\rm f/b}$ being the total rates for forward/backward
transfer of charge. 
 In turn, after exploiting the detailed balance relation $\Gamma^{\rm b}(V)=e^{-\beta eV}
 \Gamma^{\rm f}(V)$,
the linear conductance
 follows as   $G=\lim_{V\rightarrow 0}e^2\beta \Gamma^{\rm f}$. Here, 
 $\beta=1/(k_{\rm B} T)$ is the inverse thermal energy.

In general, the conditional probability $P({\mathbf N}_{\rm f},t;
{\mathbf N}_{\rm i},t_0)$ can be evaluated exactly in terms of a
multiple real time path integral involving forward $N_a(\tau)$
($n_a(\tau))$ and backward $N'_a(\tau')$ $(n'_a(\tau'))$ paths
\cite{Sassetti95}.
The effect of the bath
modes $\vartheta_a(x\ne \pm x_0/2,\tau)$ enters via
the influence functionals
 ${\cal F_{+}}=\exp\{\sum_a \Phi_+ [N_a,N'_a]\}$, and
 ${\cal F_{-}}=\exp\{\sum_a \Phi_-[n_a,n'_a]+\Phi_{\rm C}[n_a,n'_a]\}$,
   where the phases
\begin{eqnarray}
\Phi_\pm[q,q']&=&\int_{t_0}^t d\tau\int_{t_0}^t d\tau'[\dot q(\tau)-\dot q'(\tau')]\nonumber\\
&&\times \{W_{\pm}^a(\tau-\tau')\dot q(\tau')-W_{\pm}^{a*}(\tau-\tau')\dot q'(\tau')\}
\;\;\end{eqnarray}
induce nonlocal-in-time correlations
   among different tunneling transitions.
The correlation functions read \cite{Weiss99}
\begin{equation}
W^a_{\pm}(t)=\int_0^{\infty} \! \!  d\omega
\frac{J^a_{\pm} (\omega)}{\omega^2} \left[
(1-\cos \omega t ) \coth \frac{\hbar \beta \omega}{2} + i \sin \omega t \right]\;,
\label{discrcf}
\end{equation}
with the spectral densities \cite{Sassetti95}
$J_\pm^a(\omega)=\omega e^{-\omega/\omega_c}[1/2 + (\Delta E_{\sigma} /\hbar )
 \sum_{m=1}^\infty \delta(\omega -\Omega_{\pm}(m) )]/4$
for the neutral sectors ($a=\sigma,\Delta \sigma, \Delta \rho$).
 Here, we have introduced a cut-off frequency $\omega_c$, and 
 $\Omega_+(m)=\Delta E_{\sigma}(2m-1)
 /\hbar$,
 $\Omega_-=\Delta E_{\sigma}2m/\hbar$. Finally, $\Delta E_{\sigma}=\pi\hbar v_{\rm F}/x_0$ is
related to the energy quantization of the neutral plasmon modes in the
dot. In \cite{Postma01}, $\Delta E_{\sigma}> k_{\rm B}T$ also at
room temperature.  For the charged sector the spectral density reads 
 $J_{\pm}^\rho(\omega)=-K^\pm(\omega_n\to -i\omega+0^+)/\pi\hbar$, where
\begin{equation}
[K^\pm(\omega_n)]^{-1}=\frac{4v_{\rm F}}{\hbar\pi}\int_0^\infty
dk \frac{1\pm \cos(kx_0)}{\omega_n^2+\omega^2(k)}\;,
\end{equation}
and $\omega_n$ are the Matsubara frequencies.
 For an arbitrary finite range potential the spectral densities $J_\pm^\rho(\omega)$
 cannot be evaluated analytically.
In general, as in the zero-range case, $V_{\rm zr}(x)=V_0 \delta (x)$,
the leads contribute an Ohmic spectrum
at low frequency: $\lim_{\omega \to 0}J^\rho_+(\omega)+J^\rho_-(\omega)
:=J^\rho_\Sigma(\omega)=\omega/g_\rho$, where
 $g_\rho^{-2}=1+\hat{V}(k=0)/\pi\hbar v_{\rm F}$.
Theoretical estimates as well as experiments show that typical
values for $g_{\rho}$ range between 0.18 and 0.26 in SWNTs
 \cite{Egger97,Kane97,Yao99,Bockrath99}, being well below
$g_{\rho}=1$ for noninteracting electrons.

\begin{figure}
\begin{center}
\epsfig{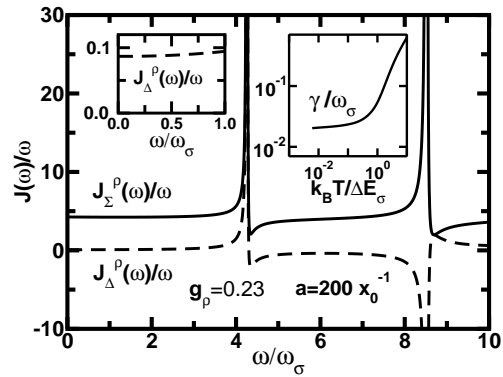}
\caption{
Spectral densities $J_{\Sigma/\Delta}^\rho(\omega)/\omega$ for
finite interaction range ($a=200
x_0^{-1}, g_\rho=0.23$) with $\omega_\sigma \equiv \Delta E_\sigma / \hbar$. 
Left inset: Enlargement of the low frequency part.  
The important parameter is $1/g_\Delta= \lim_{\omega \rightarrow 0}
 J_{\Delta}^\rho(\omega)/\omega$. 
Right inset: Contributions to the effective line-width $\Gamma^{\rm tot}$ and $\gamma^\Delta$,
 with the latter being originated from the finite range of the interaction (see text).
\label{fig.finite}}
\end{center}
\end{figure}

As a consequence of the finite range interaction, a new energy
scale arises which introduces two qualitatively new features: (i)
The $\delta$-shaped resonances for the zero-range limit turn into
resonances of finite height 
and with an intrinsic line-width $\propto 1/g_{\Delta}$, 
cf.\ Fig.\  \ref{fig.finite}. (ii)
The new  parameter $g_\Delta^{-1}$ is defined from the low
energy relation
\begin{equation}
\lim_{\omega\to 0}J_+^\rho(\omega)-J_-^\rho(\omega):=
J_\Delta^\rho(\omega)= \omega/g_\Delta\;. \label{specdelta}
\end{equation}
 For ease of calculation, we assume  a potential
   $V(x)= V_{0}a/2 \exp (-a |x|)$ \cite{Cuniberti98}, with $a$ being the inverse screening
 length ($a\rightarrow \infty$ yields the zero-range interaction
 limit $ V_0 \delta(x)$ and implies $g_\Delta^{-1}\to 0$).
 The numerical results for this  potential are shown in
 Fig.\ \ref{fig.finite}.
Note however that, as shown below, our main conclusions
are independent of the precise shape of the potential.
 Finally,  $\Phi_{\rm C}$ provides a quadratic local-in-time
 contribution proportional to
 the addition energies $E_{\sigma}=\Delta E_{\sigma}/2$
 for the neutral sectors, and $E_\rho=K^-(\omega_n=0)$ for the charge plasmons.

\begin{figure}
\begin{center}
\hfill
\epsfig{figure=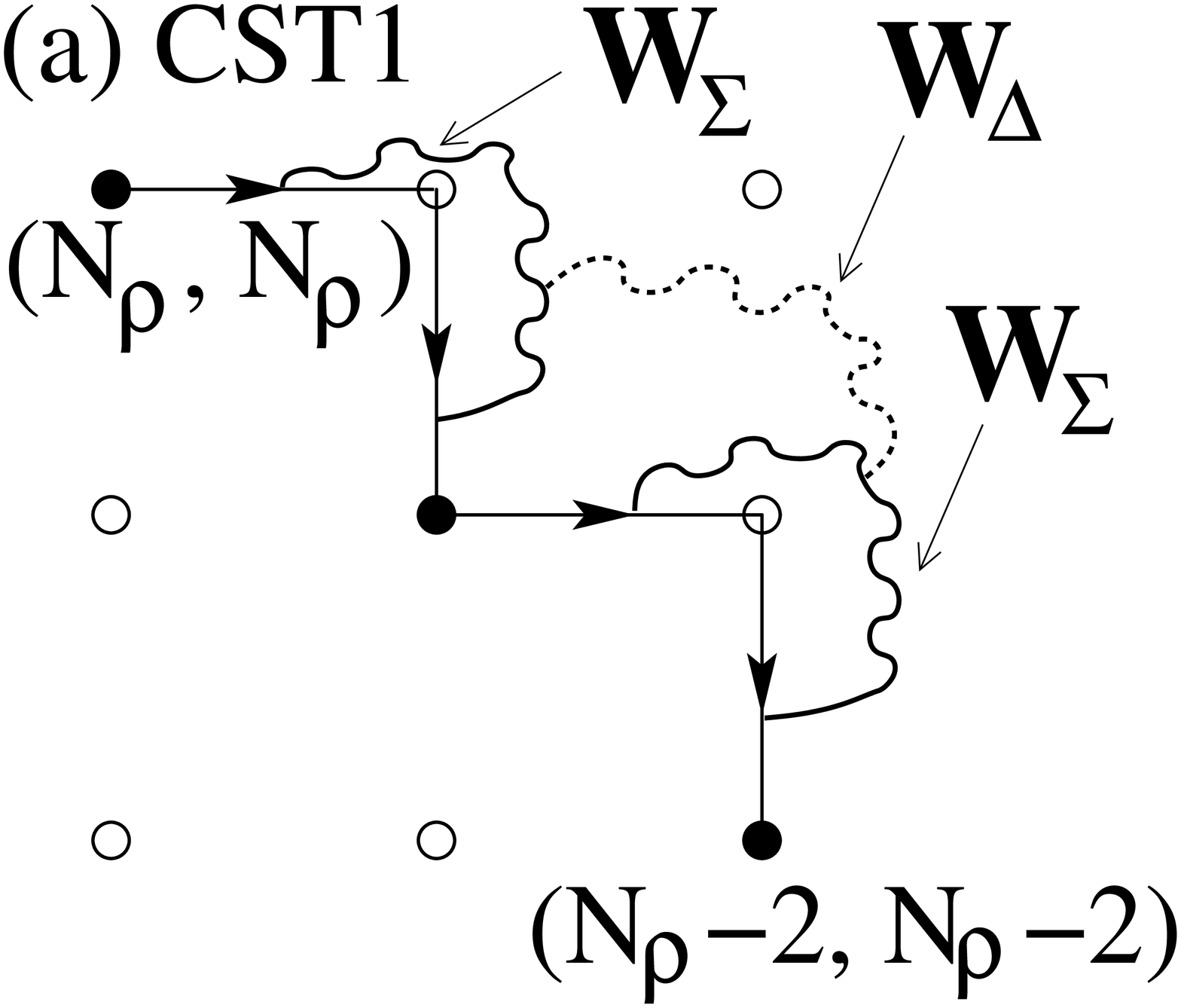,width=30mm,keepaspectratio=true}
\hfill
\epsfig{figure=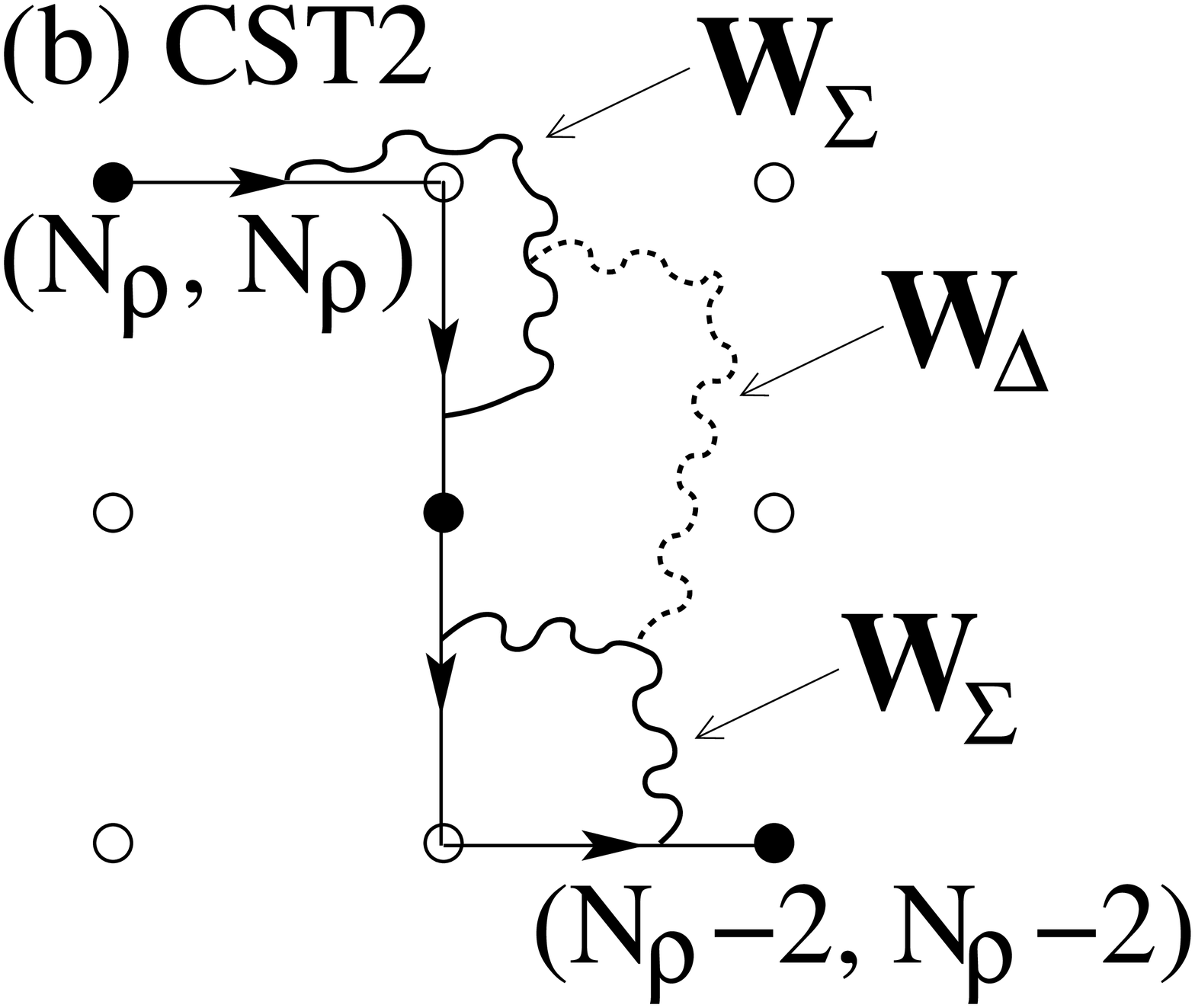,width=30mm,keepaspectratio=true}
\hfill
\caption{Lowest order
  correlated sequential tunneling (CST) paths in the $(N_\rho,N'_\rho)$-plane for the
 transfer of one  charge through the island corresponding to Eq.\ (\ref{turate}). 
 Filled circles denote diagonal states of the reduced density matrix. 
\label{fig.paths}}
\end{center}
\end{figure}

 Of foremost importance is how to
evaluate the forward rate $\Gamma^{\rm f}$. In the large barrier
limit, the dynamics is ruled by well separated tunneling events
between the adjacent minima of the periodic
 potential in $H_{\rm B}$. Each tunneling event induces the
 change $N_a\to N_a\pm 1$, and $n_a\to n_a \pm 1$.
 In particular, $N_\rho \to N_{\rho}\pm 1$ describes one discrete
 charge tunneling through
 one barrier from left to right ($-$) or vice versa ($+$),
 while $n_\rho \to n_\rho \pm 1$ describes
 a unit charge tunneling onto/out of the dot.
Each tunneling event through the left (right) barrier contributes
a factor $i\Delta_{\rm L}
(-i\Delta_{\rm R})$, where $\Delta_{\rm L}=\Delta_{\rm R}$ is the 
associated tunneling amplitude. 
 To proceed, we express the double path
integrals over the paths $N_a,N'_a$ and
 $n_a,n'_a$ as  single path integrals over paths along the states
 of the RDM in the $(N_a,N_a')$- and in the
$(n_a,n_a')$-planes.  According to the boundary conditions, all
paths start and end in diagonal states of the RDM, making
intermittent visits to off-diagonal states, cf.\ Fig.\ \ref{fig.paths}.
Due to the strong Coulomb interaction  ($g_\rho <1$),
the relevant paths are those
for which the system is back to a diagonal state after two
tunneling events \cite{Weiss99}. We focus on the linear
regime ($V\to 0$), so that either $n_\rho=n$ or $n_\rho=n+1$ units
of charge are allowed in the dot in the stationary state \cite{Kane92}.
The current (\ref{current}) is then expressed
as a series expansion in the  tunneling amplitudes $\Delta_{\rm R/L}$.
To lowest order, the forward rate is divergent. The divergent term reads 
\begin{eqnarray}
\lefteqn{\Gamma^{{\rm f},(4)}  = \lim_{\lambda\to 0}
\frac{\Delta_{\rm L}^2\Delta_{\rm R}^2}{8}
\int_0^{\infty} \left(\prod_{i=1}^3 d\tau_i\right)
 e^{-\lambda \tau_2 -S_{\Sigma}(\tau_1)-S_{\Sigma}(\tau_3)}}\nonumber\\
&&\!\!\!\!\!\!\times \cosh \Lambda_{13}^\Delta\cos {\cal R}^\Delta_{13}
[c_{\rm Lf}(\tau_1)c_{\rm Rf}(\tau_3)+c_{\rm Rf}(\tau_1)c_{\rm Lf}(\tau_3)],     
\;\;\label{turate}
\end{eqnarray}
where 
 $\tau_i$ denote the time intervals elapsed between the different
tunneling events. Then, $\tau_2$ is the time spent in the intermediate
diagonal  state. We use the notation
 $c_{\kappa\alpha}(\tau)=\cos(R_\Sigma(\tau)-E_{\kappa,\alpha}\tau )$,
where the energies  are defined as
$E_{{\rm L}, \rm f}(n)=-\mu(n+1)-eV/2$ and $E_{{\rm R}, \rm f}(n+1)=\mu(n+1)-eV/2$.
 Here,  the chemical potential reads
 $\mu(n+1)=E_{\rho}(n-n_0-C_{\rm G}V_{\rm G}/e+1/2)+(3/2)E_\sigma (\sigma_{n+1}^2-\sigma_n^2)$,
 with
$n_0=4 k_{\rm F}x_0/\pi$ being the mean electron number on the dot.
 Moreover, $\sigma_n=0$
 ($\pm 1$) for  an
 even (odd) number of spins in the dot.
The correlations are encapsulated in
 $S_{\Sigma/\Delta}(\tau)$ and $R_{\Sigma/\Delta}(\tau)$
being the real and imaginary parts, respectively, of 
 $W_{\Sigma/\Delta} \equiv \sum_a (W^a_+ \pm W^a_-)$.
The long range nature of the Coulomb interaction is reflected in the
 correlators $\Lambda^\Delta_{13}$, and ${\cal R}^\Delta_{13}$,
 inducing dipole-dipole correlations  between the tunneling events across the left and right
 barrier,  cf.\ Fig.\ \ref{fig.paths}. To be definite,
$\Lambda^{\alpha}_{13} =
S_{\alpha}(\tau_1+\tau_2+\tau_3)+S_{\alpha}(\tau_2) \
-S_{\alpha}(\tau_1+\tau_2)-S_{\alpha}(\tau_2+\tau_3)$, and ${\cal
R}^\alpha_{13}=
R_{\alpha}(\tau_1+\tau_2+\tau_3)-R_{\alpha}(\tau_2+\tau_3)+
R_{\alpha}(\tau_1+\tau_2)-R_{\alpha}(\tau_2)$. Here,
$\alpha=\Delta$ if the interaction is between 
tunneling through different barriers, while $\alpha=\Sigma$ if
 events through the same barrier are involved.
  Due to (\ref{specdelta}),  the correlators 
  associated to $W^\rho_\Delta$ 
 behave as $S_\Delta^\rho(\tau)\to \pi k_{\rm B}T\tau/g_\Delta\hbar$, and
 $R_\Delta^\rho\to$  const. at long times. Hence, $\Lambda_{13}^{\Delta,\rho}
 \to 0$
 and ${\cal R}^{\Delta,\rho}_{13}\to 0$ when
 $\tau_2\to \infty$. The functions
 $S_\Delta^a$ and $R_\Delta^a$ associated to the neutral sectors
 are purely periodic with frequency $\Delta E_\sigma/\hbar$.

To cure the divergence of the fourth order rate, we  have to sum up
  divergent irreducible  terms of  higher order.
 Therefore, appropriate approximation
schemes are required. 
 We make the following approximations: (i) We only consider higher order
paths which yield the UST result for vanishing interaction range. 
(ii) We neglect the dipole-dipole correlations $\Lambda^\Sigma_{ij}$ due to $g_\rho\ll 1$,
but keep the correlations $\Lambda^\Delta_{ij}$ among the inner dipoles up to
linear order.
 iii) We consider a large cut-off $\omega_c$ (scaling limit), and neglect
  the correlations ${\cal R}^{\Delta,\rho}_{ij}$.
The sum over the so obtained higher order irreducible paths
  yields for the total rate $\Gamma^{\rm f}=\sum_{n=2}^{\infty}
 \Gamma^{{\rm f},(2n)}$ a finite result. It can be obtained from (\ref{turate})
 upon replacing $\Lambda^\Delta_{13}\to 3\Lambda_{13}^{\Delta,\sigma}$, 
 ${\cal R}^{\Delta,\rho}_{13}=0$, and 
$\lambda \to \lambda+ \gamma$. 
  Here $\gamma$ can be interpreted as a linewidth, i.e.,
 $\gamma=\Gamma^{\rm tot}+\gamma^{\Delta}$,
 where  $\Gamma^{\rm tot}=\Gamma^{\rm Rf} + \Gamma^{\rm Lf} +\Gamma^{\rm Rb}+ \Gamma^{\rm Lb}$,
 and $\Gamma^{\kappa\alpha}=(\Delta^2_\kappa/2)\int_0^\infty
 d\tau e^{-S_\Sigma(\tau)}c_{\kappa\alpha}(\tau)$
  are the incoherent
 rates for forward/backward tunneling through the right/left barrier 
 ($E_{{\rm L}, \rm f}(n)=-E_{{\rm L}, \rm b}(n+1), 
 E_{{\rm R}, \rm b}(n)=-E_{{\rm R}, \rm f}(n+1)$).
In contrast, $\gamma^\Delta$ is the lowest
 order correction to the linewidth due to the finite range interaction.
It reads at resonance ($E_{\kappa,\alpha}=0$)
\begin{eqnarray}
\lefteqn{\gamma^\Delta=\frac{\Delta_{\rm R}^2\Delta^2_{\rm L}}{4}
\int_0^{\infty}\prod_{i=1}^3 d\tau_i e^{-S_{\Sigma}(\tau_1)-S_{\Sigma}(\tau_3)}
 e^{-\Gamma^{\rm tot}\tau_2}}\nonumber\\
&& \times
\sin R_\Sigma(\tau_1) \sin R_{\Sigma}(\tau_3) 
\Lambda^{\Delta,\rho}_{13}
\cosh 3\Lambda_{13}^{\Delta,\sigma}\;.
\end{eqnarray}
The incoherent rates yield at resonance  to the power-law
 $\Gamma^{\kappa,\alpha}\propto T^{\alpha_{\rm end}}$,
where $\alpha_{\rm end}=1/g_{\rm eff}-1$, and $1/g_{\rm
eff}=(1/g_\rho +3)/4$.
 In the limit $\Lambda_{13}^{\Delta,\rho}=0$ 
  one recovers the known UST rate   \cite{Furusaki98,Braggio00}
 $\Gamma^{\rm f} = \Gamma^{\rm Rf} \Gamma^{\rm Lf} / \Gamma^{\rm tot}
\propto T^{\alpha_{\rm end}}$, which
implies the power law (\ref{conductanceF}) for 
$G_{\rm max}$.
While $\Gamma^{\rm tot}$  decreases with decreasing  temperature,
 $\gamma^\Delta$ remains constant.
 Hence, depending on $g_{\rm eff}$ a cross-over temperature $T^*$ can be
 identified, with
 $k_{\rm B} T^* / \hbar \omega_c = (\gamma^{\Delta} / 
 d_{g_{\rm eff}})^{1-1/g_{\rm eff}}/2 \pi$
 with  $d_{g_{\rm eff}}=\Delta^2 |\Gamma(1/2 g_{\rm eff})|^2/(\omega_c\Gamma(1/g_{\rm eff}))$.
 For $T<T^*$, the power-law (\ref{conductance}) follows.
Due to the nonlinearity of the spectral densities 
$J^\rho_{\Sigma/\Delta} (\omega)$, we are able to compute the triple integral 
numerically only for rather large inverse screening lengths $a$, since 
then the correlators $W_{\Sigma/\Delta}^\rho(\tau)$ are well 
fitted by those of a damped harmonic oscillator. 
 The result for $a=200x_0^{-1}$, yielding $g_\Delta^{-1}=0.087$, is shown 
 in the right inset of Fig.\ 1 
 for  $g_\rho=0.23$. 
Turning to the experiment in \cite{Postma00} we should  a tyical 
 screening length $a^{-1}\approx  3-5$nm. For a dot length $x_0=25$nm 
 it corresponds to values $a\simeq 10 x_0^{-1}$. 
However, due to the strong asymmetry of the peaks of  
   the spectral densities $J_{\Sigma/\Delta}^\rho$
 at  $a\simeq 10 x_0^{-1}$,  a numerical evaluation of 
 $\gamma^\Delta$ is no longer possible.  In the following, we 
 assume $a= 200 x_0^{-1}$.

\begin{figure}
\begin{center}
\epsfig{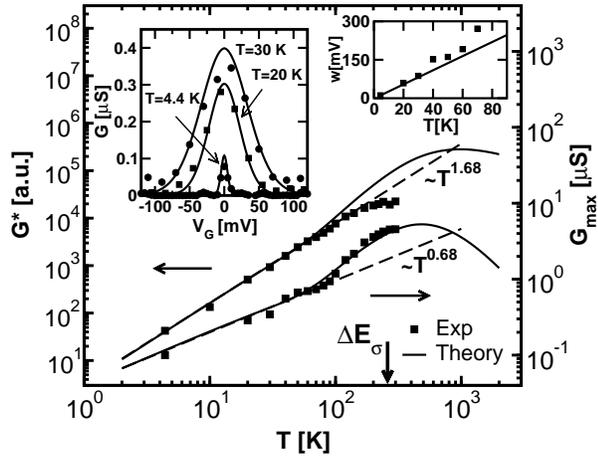}
\caption{Conductance maximum $G_{\rm max}$ (right)
and integrated conductance $G^*$ (left)
as a function of temperature.  Solid line: theoretical prediction
 with $\Delta E_\sigma=23$ meV and   interaction strength 
 $g_\rho=0.23$.
 $\blacksquare$: experimental
 results [7], dashed lines: power laws, see Eq.\  (\ref{conductance}).
  Left inset: A single conductance peak for varying gate voltage for
 $T=$4.4 K, 20 K, 30 K. Right inset:
 Peak width versus temperature.  \label{fig.powerlaw}}
 \end{center}
\end{figure}

We now compare the outcomes of (\ref{turate})
 with the experimental results  \cite{Postma01}. 
Induced buckles on a SWNT 
define an island of $\sim 25$ nm length.   
We include the contact influence via the Landauer formula,
in which 
the total tube conductance is convoluted
 with the
Fermi distribution from the Au-leads.
With typical Luttinger parameters  in the range $g_{\rho}\approx 0.18-0.26$,
the agreement between the measured data and the theoretical
 prediction 
 is very good upon choosing
 $g_\rho=0.23$,  leading to $g_{\rm eff}=0.54$.
 The comparison is depicted in Fig.\ \ref{fig.powerlaw} together with the
 power law (\ref{conductance}). Note that
 cotunneling events, which should  dominate the tails of a conductance peak
 \cite{Furusaki93,Sassetti95,Averin90},  are not considered
 here.
 The absolute value of
 $G_{\rm max}$ is fitted by the parameter $\Delta_{\rm L/R}$ (we have chosen
  $\hbar \omega_{\rm c} = 9 \Delta E_{\sigma}$).
 Importantly, we find a very good agreement between theoretical and experimental results for
  $k_{\rm B} T \ll \Delta E_{\sigma}$ 
  with the only one relevant free parameter
 $g_\rho$. 

 In the regime
 $k_{\rm B} T \simeq \Delta E_{\sigma}$, we observe a deviation from the power law
  due to the contribution of excited plasmon levels
 which  become thermally accessible, as well as due to a crossover to the UST
  limit. 
  The  result of our simple model with two $\delta-$shaped
tunneling barriers are fitted to the data. We find
 $x_0=71$ nm ($\Delta E_{\sigma}=23$ meV) which agrees with the measured
  value of  $x_0=25$ nm. A smaller value for $x_0$ only
 shifts the onset of the deviation from the power-law to higher temperatures.  Indeed,
 the measured height profile of the buckle region  suggests that
 the real scattering potential is more complicated.
   However, our central result,
namely the algebraic behavior at low temperature, is  {\em not\/} affected by the precise
shape of the impurity potentials.
We also fit the theoretically predicted line-shape
of a conductance peak to the measured data as displayed in the left
inset of Fig.\ \ref{fig.powerlaw} 
and find $\lambda= 0.25$, which
coincides with the experimentally determined value.
 We measure the width of a calculated conductance peak and
find that it
increases linearly with increasing temperature, in agreement with the measured data
(right inset of Fig.\  \ref{fig.powerlaw}).
To underpin the algebraic behavior, we also show in Fig.\  \ref{fig.powerlaw}
the conductance $G^*$  obtained by integrating the linear conductance over the gate voltage.
Due to the linear increase of the peak width with temperature,
the exponent of the power law is increased
by one, in clear agreement with the experiment.

In summary, we have shown that  the  finite range of Coulomb interaction  
can play a crucial role  in 
 the  resonant tunneling in one-dimensional quantum dots.

We thank U.\ Weiss and Yu.\ Nazarov
 for discussions and T.\ Teepen and Z.\ Yao for
experimental assistance.
We  acknowledge support  by the Dutch FOM, the 
German DFG-Emmy-Noether Programm (M.T.),  
the EC program SATURN, and the Schloe{\ss}mann Foundation (G.C.). 
%

\end{document}